\documentclass[prd,twocolumn,preprintnumbers,superscriptaddress,nofootinbib]{revtex4}
\usepackage[a4paper, hdivide={1.91cm,,1.165cm}, vdivide={1.83cm,,2.6cm}]{geometry}

\usepackage{amstext,amssymb}
\usepackage{amsmath}
\usepackage{graphicx}
\usepackage[hyperfootnotes=false]{hyperref}
\usepackage{xspace}
\usepackage{color}
\usepackage{units}
\usepackage{slashed} 

\global\long\def\d{\partial}

\begin{document}

\title{Threshold effects on prediction for proton decay in 
non-supersymmetric $E_6$ GUT with intermediate trinification symmetry}


\author{Chandini  \surname{Dash}}
\email{dash25chandini@gmail.com}
\affiliation{Department of Physics, Berhampur University, Odisha-760007, India.}


\author{Snigdha  \surname{Mishra}}
\email{mishrasnigdha60@gmail.com}
\affiliation{Department of Physics, Berhampur University, Odisha-760007, India.}

\author{Sudhanwa \surname{Patra}}
\email{sudhanwa@iitbhilai.ac.in}
\affiliation{Department of Physics, Indian Institute of Technology Bhilai, Raipur-492015, India}

\author{Purushottam \surname{Sahu}}
\email{purushottams@iitbhilai.ac.in}
\affiliation{Department of Physics, Indian Institute of Technology Bhilai, Raipur-492015, India}
\begin{abstract}

We consider a non-supersymmetric $E_6$ Grand Unified Theory (GUT) with intermediate trinification symmetry $SU(3)_C \times SU(3)_L \times SU(3)_R \times D$ (D denoted as D-parity for discrete left-right symmetry) and study the effect of one-loop threshold corrections arising due to every class of superheavy particles (scalars, fermions and vectors). It is observed that, the intermediate mass scale $M_I$ and $\sin^2\theta_W$ remain unaffected by GUT threshold contributions. The threshold modified unification mass scale $M_U$ is in excellent agreement with the present experimental proton decay constraint. The novel feature of the model is that GUT threshold uncertainty of $M_U$ is found to be controlled by superheavy scalars only, leading to a very predictive scenario for proton decay, which can be verifiable within the foreseeable experiments.

 
\vspace*{-1.0cm}
\end{abstract}


\maketitle

\section{Introduction}
\label{sec:intro}
The Standard Model(SM) of particle physics has come off with flying colors continuously with the spectacular discovery of Higgs boson at the Large Hadron Collider (LHC) in the recent past\cite{Chatrchyan:2012ufa,Aad:2012tfa,Aad:2015zhl}. Even after its trail of successes, it is now presumed that this beautiful theory of elementary particles might be an effective low energy approximation of some large Grand Unified Theory(GUT) or part of some other theory operative at a high scale. However it is observed that all GUTs, without supersymmetry(SUSY) and without an intermediate symmetry, fail to unify the three gauge couplings of the SM. With one or more intermediate symmetries, although the gauge unification is possible, it may not always comply with the present proton decay constraint which is believed to be a key prediction of most GUTs. Thus, in order to ensure favourable unification, one has to consider the possibilities, i.e. either by introducing supersymmetry\cite{Peskin:2008nw} in the GUT model or by modifying the coupling constants through non-renomalisable opetators\cite{Shafi:1983gz,Hill:1983xh} (including gravitational correction) or through the threshold effects\cite{Hall:1980kf,Mohapatra:1992jw,Parida:2016hln,Chakrabortty:2019fov,Babu:2015bna,Schwichtenberg:2018cka} at the symmetry breaking scale. However with the non-observation of the supersymmetric particles in ongoing experiments, there is an urge for non-supersymmetric GUTs like $SU(5)$~\cite{Georgi:1974sy}, $SO(10)$\cite{Pati:1974yy,Fritzsch:1974nn}, $E_6$\cite{Gursey:1975ki,Shafi:1978gg,Stech:2003sb} etc. Although the second option through gravitational effect is viable, but in the absence of specific knowledge about it's origin in the GUT model, it is tempting to revive possible non-supersymmetric GUTs with threshold effects for predictions of proton decay and other phenomenological output.

In the present paper, we consider threshold effects in a non-SUSY Grand Unified $E_6$ model with intermediate trinification symmetry $SU(3)_{C}\times SU(3)_{L}\times SU(3)_{R}$\cite{Stech:2003sb,Hetzel:2015cca} invoked with D-parity\cite{Chang:1983fu,Chang:1984uy}, unlike the conventional GUT models, where for simplicity, the superheavy fields are assumed to be exactly degenerate with the symmetry breaking scales. It is observed that the additional particles like exotic color fermions, vector-like lepton doublets and two neutral fermions contained in the fundamental representation of $E_6$ GUT along with the threshold corrections, ensure successful gauge coupling unification, in tune with the present experimental limit of proton decay lifetime. 

The paper is organized as follows. The next section is devoted to the model building along with the gauge coupling evolutions with appropriate threshold corrections. In section-III, numerical estimation of the mass scales $M_I$, $M_U$ and the GUT coupling constant $\alpha_G$ are done with specific choice of threshold parameters ensuring successful gauge unification. The successive section is devoted to numerical prediction on proton decay lifetime with threshold effects. The last section is devoted to concluding remarks on the phenomenological viability of the model.

\section{The Model Framework}
\label{sec:model}
We briefly discuss here the non-supersymmetric $E_6$ Grand Unified Theory (GUT) with one intermediate trinification symmetry $SU(3)_C\otimes SU(3)_L\otimes SU(3)_R$. Due to the presence of $SU(N)_L\otimes SU(N)_R$ (here $N=3$ for trinification symmetry) structure, there are possibility of two different scenarios of symmetry breaking chain--one with D-parity conserved and other with D-parity broken. Here we focus only on the trinification symmetry with D-parity, given as,
\begin{eqnarray}
&&\pmb{E_6} \stackrel{M_U}{\longrightarrow} SU(3)_C\otimes SU(3)_L\otimes SU(3)_R\otimes D (\mathbb{G}_{333D})\nonumber \\ 
	&&\stackrel{M_I}{\longrightarrow}SU(3)_C\otimes SU(2)_L\otimes U(1)_{Y} (\mathbb{G}_{321})\nonumber\\
	&&\stackrel{M_Z}{\longrightarrow} SU(3)_C\otimes U(1)_{Q} (\mathbb{G}_{31})
	\label{eq:ModelD}
	\end{eqnarray}
Here D-parity stands for discrete left-right symmetry~\cite{Mohapatra:1974gc, Pati:1974yy, Senjanovic:1975rk, Mohapatra:1980yp,Chang:1983fu,Chang:1984uy} which acts mostly on Higgs fields ensuring equal coupling between $g_{3L}$ and $g_{3R}$ corresponding to $SU(3)_{L}$ and $SU(3)_{R}$ symmetry. Below this D-parity breaking scale, the asymmetry in the Higgs sector gives different contributions to the beta function of  Renormalization Group Equations(RGEs) and thereby, yields asymmetry in the gauge couplings. 

The first step of spontaneous symmetry breaking in eqn.(\ref{eq:ModelD}) from $E_6$ GUT to $G_{333D}$--is achieved by giving a GUT scale VEV to D-parity even singlet scalar $(1,1,1)$ contained in $\pmb{650_H} \subset \pmb{E_6}$ leading to $g_{3L}=g_{3R}$. The next stage of symmetry breaking i.e from $\mathbb{G}_{333D} \to \mathbb{G}_{\rm 321}$ is done by assigning a non-zero VEV to the $\mathbb{G}_{\rm 321}$ neutral component of trinification multiplet $(1,\overline{3},3)$ of $\pmb{27_H}$ and $(1,8,8)$ of $\pmb{650_H}$ of $\pmb{E_6}$. It has been shown in some earlier works \cite{Dash:2019bdh,Stech:2003sb,Chakrabortty:2009xm,Chakrabortty:2017mgi} that this model with minimal Higgs $(1,\overline{3},3) \subset \pmb{27_H}$ could not admit phenomenologically viable gauge unification(which comes out to be the scale beyond Planck energy). So additional Higgs of $\pmb{351_H}$ and $\pmb{351^{'}_H}$ were used to achieve the goal. However in the present work we overcome the problem by confining the Higgs sector with the multiplet $(1,8,8)\subset\pmb{650_H}$ to maintain the minimal feature of the model. The last stage of symmetry breaking i.e SM to low energy theory $(G_{31})$ is done by assigning a non-zero VEV to SM Higgs doublet contained in $\pmb{27_H} \subset \pmb{E_6}$. We follow the \textquotedblleft Extended Survival Hypothesis\textquotedblright for Higgs scalars responsible for spontaneous symmetry breaking and their contributions to RGEs by deriving one-loop beta functions along with one-loop threshold corrections in the following discussion.

%
Now in order to obtain the gauge coupling evolution we use the standard Renormalization Group Equations (RGEs)\cite{Georgi:1974yf} for different range of mass scales corresponding to the channel in eqn.(\ref{eq:ModelD}). Here we include the threshold effects both at the intermediate mass scale $M_I$ and the unification mass scale $M_U$.Due to the threshold effects, the matching condition at the symmetry breaking scale $\mu$ is modified \cite{Hall:1980kf} as,
\begin{eqnarray}
 \alpha^{-1}_{D} (\mu)&&=\alpha^{-1}_{P} (\mu)-\frac{\pmb{\lambda_{D}(\mu)}}{12\pi}
\end{eqnarray}
where $P$, the parent (simple or product) gauge group (with inverse coupling constant $\alpha_{P}^{-1}(\mu)$), is broken to the daughter (simple or product) group $D$ (with inverse coupling constant $\alpha_{D}^{-1}(\mu)$) at the mass scale $\mu$. Here $\mu=M_I$ and $M_U$ where threshold effects are considered.
The corresponding group equations are given by\\
{\bf (i) Between the mass scale $M_Z$ to $M_I$:-}\\
\begin{eqnarray}
\alpha^{-1}_{3C} (M_Z)&&=\alpha^{-1}_{3C} (M_I) + \frac{\pmb{b_{3C}}}{2 \pi} {\large \ln}\left(\frac{M_I}{M_Z}\right)-\frac{\pmb{\lambda_{3C}^{I}}}{12\pi} \nonumber\\
\alpha^{-1}_{2L} (M_Z)&&=\alpha^{-1}_{3L} (M_I)+ \frac{\pmb{b_{2L}}}{2 \pi} {\large \ln}\left(\frac{M_I}{M_Z}\right)-\frac{\pmb{\lambda_{2L}^{I}}}{12\pi} \nonumber\\
\alpha^{-1}_{Y} (M_Z)&&=\frac{1}{5}\alpha^{-1}_{3L} (M_I)+\frac{4}{5}\alpha^{-1}_{3R} (M_I)\nonumber\\
&&+\frac{\pmb{b_{Y}}}{2 \pi} {\large \ln}\left(\frac{M_I}{M_Z}\right)-\frac{\pmb{\lambda_{Y}^{I}}}{12\pi}
\label{eqn:alphaMI}
\end{eqnarray}
where $\pmb{b_{i}}\, (\mbox{i=3C,2L,1Y})$ are one-loop beta coefficients  and $\pmb{\lambda_{i}^{I}}\, (\mbox{i=3C,2L,1Y})$ are the one-loop threshold effects arising due to the superheavy fields at the intermediate mass scale $M_I$.

{\bf (ii) Between the mass scale $M_I$ to $M_U$:-}\\
\begin{eqnarray}
\alpha^{-1}_{i} (M_I)&&=\alpha^{-1}_{G} (M_U)+ \frac{\pmb{b^\prime_{i}}}{2 \pi} {\large \ln}\left(\frac{M_U}{M_I}\right)-\frac{\pmb{\lambda_{i}^{U}}}{12\pi}
\label{eqn:alphaMU}
\end{eqnarray}
Here $\alpha^{-1}_{G}$ is the $E_6$ GUT coupling constant and $\pmb{b^\prime_{i}}\, (\mbox{i=3C,3L,3R})$ are one-loop beta coefficients. $\pmb{\lambda_{i}^{U}}\, (\mbox{i=3C,3L,3R})$ are the one-loop threshold effects at the unification mass scale $M_U$. The detail expression will be discussed in the next section. \\
Now to obtain the numerical value of one-loop beta-coefficients $\pmb{b_i}$, we use the general expression
\begin{eqnarray}
	&&\pmb{b_i}= - \frac{11}{3} C_2(G) 
				 + \frac{2}{3} \,\sum_{R_F} T(R_F) \prod_{j \neq i} d_j(R_F) \nonumber \\
  &&\hspace*{2.5cm} + \frac{1}{3} \sum_{R_S} T(R_S) \prod_{j \neq i} d_j(R_S).
\label{oneloop_bi}
\end{eqnarray}

where the notations have their usual meanings with first term denotes gauge bosons contribution, second term arises due to fermions and third term is due to Higgs scalars.
The one-loop beta coefficients for the present model, are given in Table \ref{tab:oneloopresults}.

\begin{table}[h!]
\begin{tabular}{|c|c|c|c|c|}
\hline
Group & Range  & Higgs & Fermions & beta coefficients \\

 &  of masses & content & content  &  \\
\hline
$G_{321}$ & $M_{Z}-M_{I}$ & $ \begin{array}{clcr}
                                \phi(1,2,-\frac{1}{2})_{27} 
                                 \end{array} $ &$ \begin{array}{clcr}
                               Q(3,2,\frac{1}{6})\\ u^C(\overline{3},1,-\frac{2}{3})\\d^C(\overline{3},1,\frac{1}{3})\\l(1,2,-\frac{1}{2}) \\e^C(1,1,1)
                                 \end{array} $ &$\begin{pmatrix}\pmb{b_{3C}}=-7\\
                                 \pmb{b_{2L}}=-\frac{19}{6}\\
                                 \pmb{b_{Y}}=\frac{41}{10}\end{pmatrix}$\\
\hline
$G_{333}$ & $M_{I}-M_{U}$ & $ \begin{array}{clcr}
                              \phi_{1}(1,\overline{3},3)_{27}\\
                              \Sigma_{1}(1,8,8)_{650} \end{array} $ & $ \begin{array}{clcr}
                              L(1,\overline{3},3)\\
                              Q(3,3,1)\\ Q^C(\overline{3},1,\overline{3}) \end{array} $ &$\begin{pmatrix} \pmb{b_{3C}^{\prime}}=-5 \\ \pmb{b_{3L}^{\prime}}=\frac{7}{2} \\ \pmb{b_{3R}^{\prime}}=\frac{7}{2} \end{pmatrix}$\\
\hline
\end{tabular}
\caption{ Higgs fields and one-loop beta coefficients 
for different range of masses}.
\label{tab:oneloopresults}
\end{table}
Now using the evolution equations (\ref{eqn:alphaMI}) and (\ref{eqn:alphaMU}) for the gauge couplings, we obtain the following relations:
\begin{eqnarray}
&&\alpha^{-1}_{3C} (M_Z)=\alpha^{-1}_{G} + \frac{\pmb{b_{3C}}}{2 \pi} {\large \ln}\left(\frac{M_I}{M_Z}\right) + \frac{\pmb{b^\prime_{3C}}}{2 \pi} {\large \ln} \left(\frac{M_U}{M_I}\right)
\nonumber \\
&&\hspace*{3cm}-\frac{\pmb{\lambda_{3C}^{I}}}{12\pi} -\frac{\pmb{\lambda_{3C}^{U}}}{12\pi} 
\label{eqn:3C-I}
   \end{eqnarray}
 \begin{eqnarray}
&&\alpha^{-1}_{2L} (M_Z)=\alpha^{-1}_{G} + \frac{\pmb{b_{2L}}}{2 \pi} {\large \ln}\left(\frac{M_I}{M_Z}\right) + \frac{\pmb{b^\prime_{3L}}}{2 \pi} {\large \ln} \left(\frac{M_U}{M_I}\right)
\nonumber \\
&&\hspace*{3cm}-\frac{\pmb{\lambda_{2L}^{I}}}{12\pi} -\frac{\pmb{\lambda_{3L}^{U}}}{12\pi}   \label{eqn:2L-I}     
\end{eqnarray}
 \begin{eqnarray}
&&\alpha^{-1}_{Y} (M_Z)=\alpha^{-1}_{G}  + \frac{\pmb{b_Y}}{2 \pi} {\large \ln}\left(\frac{M_I}{M_Z}\right)+ \frac{\frac{1}{5} \pmb{b^\prime_{3L}} 
 + \frac{4}{5} \pmb{b^\prime_{3R}}}{2 \pi} {\large \ln} \left(\frac{M_U}{M_I}\right) 
 \nonumber \\
&&\hspace*{3cm}-\frac{\pmb{\lambda_{Y}^{I}}}{12\pi}-\left(\frac{\frac{1}{5} \pmb{\lambda_{3L}^{U}}  + \frac{4}{5} \pmb{\lambda_{3R}^{U}}}{12 \pi} \right)
 \label{eqn:Y-I}
\end{eqnarray}
 In order to obtain the analytical expression for the intermediate mass scale $M_I$, the unification mass scale $M_U$, the inverse GUT coupling constant $\alpha_G^{-1}$ and the electroweak mixing angle $\sin^2\theta_W$, we  use the standard key relations, $\alpha^{-1}_{em}(M_Z)-\frac{8}{3}\alpha^{-1}_{3C} (M_Z)$, $\alpha^{-1}_{em}(M_Z)-\frac{8}{3}\alpha^{-1}_{2L} (M_Z)$ and $\alpha^{-1}_{em}(M_Z)=\frac{5}{3}\alpha^{-1}_{Y} (M_Z)+\alpha^{-1}_{2L}(M_Z)$.\\

 These are given as,
\vspace*{0.15cm}
\begin{eqnarray}
&&{\large \ln}\left(\frac{M_U}{M_Z}\right) = 
 \frac{A_{I} \pmb{D_{W}}-B_{I} \pmb{D_{S}}}{B_{U} A_{I}-B_{I} A_{U}} + \frac{B_{I} \pmb{J_{\lambda}}-A_{I} \pmb{K_{\lambda}}}{B_{U} A_{I}-B_{I} A_{U}} \nonumber \\
 &&\hspace*{1.65cm}={\large \ln}\left(\frac{M_U}{M_Z}\right)_{1-loop}+\pmb{\Delta}{\large \ln}\left(\frac{M_U}{M_Z}\right)_{Threshold}
 \label{rel:MU}
\end{eqnarray}
\begin{eqnarray}
\alpha_{G}^{-1} &=&
 \frac{3}{8}\Bigg[\alpha^{-1}_{\rm em}(M_Z)-\frac{C_I}{2\pi}{\large \ln}\left(\frac{M_I}{M_Z}\right)-\frac{C_U}{2\pi}{\large \ln}\left(\frac{M_U}{M_Z}\right)\Bigg]_{1-loop}
 \nonumber \\
 &&\hspace*{-0.80cm}+\frac{3}{8}\Bigg[-\frac{C_I}{2\pi}\pmb{\Delta}{\large \ln}\left(\frac{M_I}{M_Z}\right)-\frac{C_U}{2\pi}\pmb{\Delta}{\large \ln}\left(\frac{M_U}{M_Z}\right)+\pmb{F_{\lambda}}\Bigg]_{Threshold} \nonumber \\
&=&(\alpha_{G}^{-1})_{1-loop}+\pmb{\Delta} (\alpha_{G}^{-1})_{Threshold}
 \label{rel:alphaG}
\end{eqnarray}
\begin{eqnarray}
 &&{\large \ln}\left(\frac{M_I}{M_Z}\right)=\frac{B_{U} \pmb{D_{S}}-A_{U} \pmb{D_{W}}}{B_{U} A_{I}-B_{I} A_{U}} + \frac{A_{U} \pmb{K_{\lambda}}-B_{U} \pmb{J_{\lambda}}}{B_{U} A_{I}-B_{I} A_{U}} \nonumber\\
 &&\hspace*{1.65cm}={\large \ln}\left(\frac{M_I}{M_Z}\right)_{1-loop}+\pmb{\Delta}{\large \ln}\left(\frac{M_I}{M_Z}\right)_{Threshold}
 \label{rel:MI}
\end{eqnarray}
\begin{eqnarray}
 \sin^2\theta_W&=& \frac{1}{A_U}\Big[\frac{3}{8}A_U 
 +\left(\frac{\alpha_{\rm em}}{\alpha_s}-\frac{3}{8}\right)B_U 
\nonumber \\
&+& \frac{\alpha_{\rm em}\left(A_{U}B_{I}-A_{I}B_{U}\right)}{16\pi}\ln\left(\frac{M_I}{M_Z}\right) \Big]_{1-loop}
\nonumber \\
&+& \frac{1}{A_U}\Big[\frac{\alpha_{\rm em}\left(A_{U} \pmb{K_{\lambda}} -B_{U} \pmb{J_{\lambda}}\right)}{16\pi} \Big]_{Threshold} \nonumber \\
&=& (\sin^2\theta_W)_{1-loop}+\pmb{\Delta} (\sin^2 \theta_W)_{Threshold}
\label{rel:sinsqthetaw}
\end{eqnarray}
 The parameters $D_S$, $D_W$ are the electroweak precision datas \cite{Tanabashi:2018oca}, $A_I$, $A_U$, $B_I$, $B_U$, $C_I$ and $C_U$  correspond to the contribution of one-loop effects. Similarly, the parameters $\pmb{J_{\lambda}}$, $\pmb{K_{\lambda}}$ and $\pmb{F_{\lambda}}$ correspond to the threshold effects. Using the one-loop beta coefficients of the present model, the analytical expression for the parameters are given by,

\begin{eqnarray}
&&\pmb{D_S}= 16\pi \left[\alpha_S^{-1}(M_Z)-\frac{3}{8}\alpha^{-1}_{\rm em}(M_Z) \right]
=\pmb{-1986.18} \nonumber \\
&&\pmb{D_W}= 16\pi \alpha^{-1}_{\rm em}(M_Z)\left[\sin^2\theta_W-\frac{3}{8} \right]
=\pmb{-924.266}\\
&&A_I= \Bigg[\Big(8\pmb{b_{3C}}-3\pmb{b_{2L}}-5\pmb{b_{Y}} \Big)
-\Big(8\pmb{b_{3C}^{\prime}}-4\pmb{b_{3L}^{\prime}}-4\pmb{b_{3R}^{\prime}} \Big) \Bigg]=\pmb{1} \nonumber \\
&& A_U= 4\Big (2\pmb{b_{3C}^{\prime}}-\pmb{b_{3L}^{\prime}}-\pmb{b_{3R}^{\prime}} \Big) =\pmb{-68} \nonumber \\
&&B_I=\Big[5\Big (\pmb{b_{2L}}-\pmb{b_{Y}} \Big)-4\Big (\pmb{b_{3L}^{\prime}}-\pmb{b_{3R}^{\prime}} \Big) \Big] =\pmb{-\frac{109}{3}} \nonumber \\
&&B_U=4 \Big ( \pmb{ b^\prime_{3L} } -  \pmb{ b^\prime_{3R} } \Big)= \pmb{0} 
\label{eqn:BU} \nonumber\\
&&C_I=\left[\left (\frac{5}{3}\pmb{b_{Y}}+\pmb{b_{2L}} \right)-\left (\frac{4}{3}\pmb{b_{3L}^{\prime}}+\frac{4}{3}\pmb{b_{3R}^{\prime}} \right) \right] =\pmb{-\frac{17}{3}} \nonumber \\
&&C_U= \frac{4}{3}\Big( \pmb{ b^\prime_{3L} } + \pmb{ b^\prime_{3R} } \Big)=\pmb{\frac{28}{3}}\\
&&\pmb{J_{\lambda}} = \frac{1}{6}\Big[\Big( 5\pmb{\lambda^I_{Y}}+ 3\pmb{\lambda^I_{2L}}-8 \pmb{\lambda^I_{3C}} \Big)+\Big( 4\pmb{\lambda^U_{3L}}+ 4\pmb{\lambda^U_{3R}}-8 \pmb{\lambda^U_{3C}} \Big)\Big]
\label{eqn:lambdaJ} \nonumber \\
&&\pmb{K_{\lambda}} = \frac{1}{6}\Big[\Big( 5\pmb{\lambda^I_{Y}}-5\pmb{\lambda^I_{2L}}\Big)+\Big(4\pmb{\lambda^U_{3R}}-4\pmb{\lambda^U_{3L}} \Big)\Big]
\label{eqn:lambdaK} \nonumber \\
&&\pmb{F_{\lambda}}= \frac{1}{12\pi}\Bigg[\Big( \frac{5}{3}\pmb{\lambda^I_{Y}}+\pmb{\lambda^I_{2L}}\Big)+ \Big(\frac{4}{3}\, \pmb{\lambda^{U}_{3L}}+\frac{4}{3}\, \pmb{\lambda^{U}_{3R}}\Big)\Bigg]\,
\label{eqn:lambdaF}
\end{eqnarray}

 In the expressions of $\pmb{J_{\lambda}}$, $\pmb{K_{\lambda}}$ and $\pmb{F_{\lambda}}$, the first and second term denote the threshold effects at $M_I$ and $M_U$ respectively. With no threshold effects i.e. $\pmb{\lambda_i^{I,U}}=0$, we obtain, $M_U=10^{14.81}$ GeV, $\alpha_G^{-1}=40.1059$, $M_I=10^{13.007}$ GeV and $\sin^2\theta_W=0.23129$. \\
 \\
 With threshold effects it is noteworthy to mention a nice property of the present model, that values of the electroweak mixing angle $\sin^2\theta_W$ and the intermediate mass scale $M_I$ have vanishing contributions due to GUT threshold effects, similar to the proposition made in a recent paper \cite{Dash:2019bdh} with reference to one-loop, two-loop and gravitational correction. The analytical expression of GUT threshold contributions are given by
\begin{eqnarray}
 \pmb{\Delta}{\large \ln}\left(\frac{M_I}{M_Z}\right)_{Threshold}=\frac{\frac{2}{3} \Big(\pmb{\lambda^U_{3L}}-\pmb{\lambda^U_{3R}} \Big)}{B_{I}}\nonumber\\
 \pmb{\Delta} (\sin^2 \theta_W)_{Threshold}=\frac{\frac{2}{3}\alpha_{\rm em} \Big(\pmb{\lambda^U_{3R}}-\pmb{\lambda^U_{3L}} \Big)}{16\pi}
 \label{eqn:vanSM}
\end{eqnarray}
These uncertainties will vanish with $\pmb{\lambda_{3L}^U=\lambda_{3R}^U}$ as a result of the existing left-right discrete symmetry (conserved D-parity) of  the model. This equality of the threshold effect arises due to left-right symmetric superheavy particles (scalars, gauge bosons and fermions) present in the model. The detail expressions will be discussed in the next section. The key matching condition between the gauge couplings,
\begin{eqnarray}
 \alpha^{-1}_{Y}(M_I) = \frac{1}{5} \alpha^{-1}_{3L}(M_I) + \frac{4}{5}\alpha^{-1}_{3R}(M_I)
\end{eqnarray}
also plays a significant role \cite{Dash:2019bdh} for the vanishing of the threshold uncertainty. This proposition is independent of the choice of particle contents, hence can be generalised for both non-supersymmetric and supersymmetric version of every class of Grand Unified Theories accommodating intermediate trinification symmetry $G_{333D}$. 
\vspace*{-0.60cm}
\section{Model predictions for the mass scales $M_U$, $M_I$ and the GUT coupling constant with threshold effects}
\label{sec:predictions}
 It is known that threshold effects arise from the modification of light gauge boson propagator in the effective theory due to superheavy gauge bosons, scalars and fermions in the loop. In the present model, threshold effect $\lambda_i^I$ and $\lambda_i^U$  arises due to the superheavy fields around $M_I$ and $M_U$ respectively. The general expression for $\lambda_i(\mu)$ is given by
\begin{eqnarray}
\pmb{\lambda_i}(\mu)&=& 
\mbox{\large Tr} \left(\pmb{t_{iV}^2}\right)
  -21\, \mbox{\large Tr} \left[\pmb{t_{iV}^2} \large {\large \ln} \left(\frac{\pmb{M_V}}{\mu}\right) \right] \nonumber \\
  &&\hspace*{-1.3cm}+2k \mbox{\large Tr} \left[\pmb{t_{iS}^2} \large {\large \ln }\left(\frac{\pmb{M_S}}{\mu}\right)\right] + 8 \kappa\, \mbox{\large Tr} \left[\pmb{t_{iF}^2} \large {\large \ln} \left(\frac{\pmb{M_F}}{\mu}\right)\right]
  \label{eq:threshold}
\end{eqnarray}
In eqn.(\ref{eq:threshold}), the first two terms represent threshold effects due to superheavy gauge bosons, the third term is the threshold effects due to superheavy scalars while the fourth term accounts for threshold effects due to superheavy fermions. And $\pmb{t_{iV}}$, $\pmb{t_{iS}}$ and $\pmb{t_{iF}}$ 
are denoting generators of the superheavy vector gauge bosons, scalars and fermions, respectively, under the gauge group $\mathbb{G}_j$. Here $k=\frac{1}{2}(=1)$ for real scalar fields (for complex scalar fields) while $\kappa= \frac{1}{2}(=1)$ is for Weyl fermions (for Dirac fermions) in the last term of eqn.(\ref{eq:threshold}). The notations $M_V$, $M_S$ and $M_F$ in eqn.(\ref{eq:threshold}) are the masses of the superheavy vector gauge bosons, scalars and fermions respectively.
The superheavy fields with masses around the symmetry breaking scale contribute to the threshold corrections by using \textquotedblleft Extended Survival Hypothesis\textquotedblright.
Now using Table \ref{tab:app} in the appendix, the superheavy components(fermions, scalars and vector bosons) to be used in our calcualtion, are given as,\\
\\
{\bf (i) Superheavy particles under $G_{333D}$ at $M_U$:-}\\
\begin{eqnarray}
\pmb{27_F}&& \supset \mbox{No superheavy fermions} \nonumber \\
\pmb{27_H}&&\supset \Phi_{2}(3,3,1),\Phi_{3}(\overline{3},1,\overline{3}) \nonumber \\
\pmb{650_H}&&\supset  \Sigma_{2}(1,1,1),\Sigma_{3}(1,8,1),\Sigma_{4}(1,1,8),\Sigma_{5}(8,1,1) \nonumber \\
&&\Sigma_{6}(\overline{3},3,3),\Sigma_{7}(\overline{3},3,3),\Sigma_{8}(3,\overline{3},\overline{3}) \nonumber \\
&&\Sigma_{9}(3,\overline{3},\overline{3}),\Sigma_{10}(3,6,\overline{3}),\Sigma_{11}(3,\overline{3},6)\nonumber \\
&&\Sigma_{12}(\overline{3},\overline{6},3),\Sigma_{13}(\overline{3},3,\overline{6}),\Sigma_{14}(\overline{6},\overline{3},\overline{3}) \nonumber \\
&&\Sigma_{15}(6,3,3),\Sigma_{16}(8,1,8),\Sigma_{17}(8,8,1) \nonumber \\
\pmb{78_V}&&\supset  V_4(\overline{3},3,3), V_5(3,\overline{3},\overline{3})
\end{eqnarray}
{\bf (ii) Superheavy particles under $G_{321}$ at $M_I$:-}\\
\begin{eqnarray}
\pmb{27_F}&& \supset D(3,1,-\frac{1}{3}), D^{C}(\overline{3},1,\frac{1}{3}), \psi(1,2,-\frac{1}{2})\nonumber \\
&&\psi^{C}(1,2,\frac{1}{2}), \rho(1,1,0), \nu^{C}(1,1,0) \nonumber\\
\pmb{27_H}&&\supset \Phi_{11}(1,1,0),\Phi_{12}(1,2,-\frac{1}{2}), \Phi_{13}(1,2,\frac{1}{2}) \nonumber \\
&& \Phi_{14}(1,1,1), \Phi_{15}(1,1,0)  \nonumber\\
\pmb{78_V}&&\supset  V_{21}(1,1,0), V_{22}(1,2,\frac{1}{2}), V_{23}(1,2,-\frac{1}{2})\nonumber \\
&&V_{31}(1,1,1), V_{32}(1,1,0), V_{33}(1,1,0), V_{34}(1,1,-1)\nonumber \\
&& V_{35}(1,1,1), V_{36}(1,1,-1), V_{37}(1,1,0)
\end{eqnarray}
We now make the following assumptions for the mass parameters of the superheavy particles,\\
{\bf (i) At the GUT scale $M_U$:-}
\begin{itemize}
 \item The superheavy scalars belonging to a specific multiplet of $E_6$ have degenerate mass, i.e. scalars of $27_H$ have degenerate mass $M_{S_{1}}^U$ and scalars of $650_H$ have degenerate mass $M_{S_{2}}^U$. 
 \item The superheavy gauge bosons belonging to $78_V$ attain degenerate mass $M_V^U$. 
\item There are no superheavy fermions at the GUT scale, since all the fermions i.e. SM and the exotic fermions, contained in $Q$, $Q^C$ and $L$ of trinification symmetry belonging to $27$ of $E_6$ remain light.
\end{itemize}
{\bf (ii) At the intermediate mass scale $M_I$:-}
\begin{itemize}
\item Superheavy scalars of $27_H$ have degenerate mass $M_S^I$. 
\item The superheavy fermions $\in 27_F$ and gauge bosons $\in 78_V$ attain degenerate mass $M_F^I$ and $M_V^I$ respectively.
\end{itemize}
Following the above assumptions,
we obtain $\pmb{\lambda_{i}^{U}} (i=3C,3L,3R)$, 
\begin{eqnarray}
 \pmb{\lambda_{3C}^{U}}= 9+6\eta_{S_{1}}^{U}+300\eta_{S_{2}}^{U}-189\eta_V^U  \nonumber\\
 \pmb{\lambda_{3L}^{U}}= 9+3\eta_{S_{1}}^{U}+252\eta_{S_{2}}^{U}-189\eta_V^U \nonumber \\
 \pmb{\lambda_{3R}^{U}}= 9+3\eta_{S_{1}}^{U}+252\eta_{S_{2}}^{U}-189\eta_V^U 
 \label{eqn:LambdaU}
\end{eqnarray}
where $\eta_{S_{1}}^U=\ln \frac{M_{S_{1}}^U}{M_U}$, $\eta_{S_{2}}^U=\ln \frac{M_{S_{2}}^U}{M_U}$, $\eta_V^U=\ln \frac{M_V^U}{M_U}$. Here we note that $\pmb{\lambda_{3L}^U=\lambda_{3R}^U}$ which obviously leads to the vanishing of the GUT threshold contributions for $M_I$ and $\sin^2\theta_W$ as has been mentioned(eqn.(\ref{eqn:vanSM})) in section-\ref{sec:model}.
Similarly $\pmb{\lambda_{i}^{I}} (i=3C,2L,Y)$, are given by
\begin{eqnarray}
&&\pmb{\lambda_{3C}^I}= 8\eta_F^I\,, \quad  \pmb{\lambda_{2L}^I}=1+2\eta_S^I+8\eta_F^I-21\eta_V^I\,, \nonumber\\
&&\pmb{\lambda_{Y}^I}=3+\frac{12}{5}\eta_S^I+8\eta_F^I-63\eta_V^I
 \label{eqn:LI-Th}
\end{eqnarray}

  where $\eta_S^I= \ln \frac{M_S^{I}}{M_I}$, $\eta_F^I=\ln \frac{M_F^{I}}{M_I}$ and $\eta_V^I= \ln \frac{M_V^{I}}{M_I}$.

Using the one-loop beta coefficients(Table (\ref{tab:oneloopresults})) and the parameters $\pmb{J_{\lambda}}$, $\pmb{K_{\lambda}}$ and $\pmb{F_{\lambda}}$ (from eqns. (\ref{eqn:lambdaJ})),
we have the threshold uncertainty of the unification mass scale $M_U$, the GUT coupling constant $\alpha_G^{-1}$, the intermediate mass scale $M_I$, electroweak mixing angle $\sin^2\theta_W$ given as 

\begin{eqnarray}
&&
 \pmb{\Delta}{\large \ln}\left(\frac{M_U}{M_Z}\right) 
=\frac{1}{5559}\Big[70\lambda_Y^I+39\lambda_{2L}^I-109\lambda_{3C}^I
\nonumber \\
 &&\hspace*{3cm}+109\lambda_{3R}^U-109\lambda_{3C}^U\Big] \label{eq:LMU} \\
&&
 \pmb{\Delta} \alpha_G^{-1} 
 =\frac{1}{22236\pi}\Big[1210\lambda_Y^I-120\lambda_{2L}^I+763\lambda_{3C}^I
\nonumber \\
 &&\hspace*{3cm}+1090\lambda_{3R}^U+763\lambda_{3C}^U\Big]
 \label{eq:LalphaG}\\
 &&
 \pmb{\Delta}{\large \ln}\left(\frac{M_I}{M_Z}\right)
 =\frac{5}{218}\Big[\lambda_Y^I-\lambda_{2L}^I\Big] 
 \label{eq:LMI}\\
 &&\pmb{\Delta} \sin^2\theta_W 
 =\frac{5\alpha_{em}}{96\pi}\Big[\lambda_Y^I-\lambda_{2L}^I\Big] 
 \label{eq:Lsin}
 \end{eqnarray}
 
Using eqns.(\ref{eqn:LambdaU}) and (\ref{eqn:LI-Th}) in eqns.(\ref{eq:LMU})-(\ref{eq:Lsin}), we have
\vspace*{-0.2cm}
\begin{eqnarray}
&&
 \pmb{\Delta}{\large \ln}\left(\frac{M_U}{M_Z}\right) 
=\frac{1}{1853}\Big[83+82\eta_S^I
\nonumber \\
 &&\hspace*{2.15cm}-1743\eta_V^I-109(\eta_{S_{1}}^{U}+16\eta_{S_{2}}^{U})\Big] \label{eq:rel1} \\
&&
 \pmb{\Delta} \alpha_G^{-1} 
 =\frac{1}{22236\pi}\Big[20187+2664\eta_S^I+14824\eta_F^I 
            \nonumber \\
 &&\hspace*{-0.3cm}-73710\eta_V^I +7848\eta_{S_{1}}^{U}+503580\eta_{S_{2}}^{U}-350217\eta_V^U\Big]
 \label{eq:rel2}\\
 &&
 \pmb{\Delta}{\large \ln}\left(\frac{M_I}{M_Z}\right)
 =\frac{1}{109}\Big[5+\eta_S^I-105\eta_V^I\Big] \label{eq:rel3}\\
 &&
 \pmb{\Delta} \sin^2\theta_W 
 =\frac{\alpha_{em}}{48\pi}\Big[5+\eta_S^I-105\eta_V^I\Big] 
 \label{eq:rel4}
 \end{eqnarray}

 For viable phenomenology, we then fine tune $\eta_S^I$ and $\eta_V^I$ such that the threshold uncertainty for $\sin^2\theta_W$ is in agreement with the experimental uncertainty i.e., $\sin^2\theta_W=0.23129\pm 0.00005$ \cite{Tanabashi:2018oca}. Thus referring to eqn. (\ref{eq:rel4}), we fix $\eta_S^I$ (mass of scalar) and $\eta_V^I$ (mass of gauge boson) such that $\frac{\alpha_{em}}{48\pi}\Big[5+\eta_S^I-105\eta_V^I\Big]=\pm0.00005$ to meet the experimental uncertainty. This will also ensure the stability of the intermediate mass scale $M_I$ which is essential for prediction of neutrino masses etc. Keeping this constraint in view, we choose the masses of the other superheavy particles so as to achieve admissible gauge unification in tune with proton decay lifetime. Now referring to eqns.(\ref{eq:rel1}) and (\ref{eq:rel2}), we parameterise $\eta_{S_{1}}^{U}$ and $\eta_{S_{2}}^{U}$ in such a manner, so as to obtain $M_U$ in the range $10^{15.6}$ GeV to $10^{16}$ GeV with admissible value of $\alpha_G^{-1}$.
 The estimation of $M_I$, $M_U$, $\sin^2\theta_W$ and $\alpha_G^{-1}$ given in Table \ref{tab:threresultsMI}. 

\begin{widetext}
\begin{center}
\begin{table}[h]
\centering
\vspace{-2pt}
\begin{tabular}{||c|c|c|c|c|c|c|c|c|c||}
\hline \hline
$\pmb{\eta_S^I}$& $\pmb{\eta_V^I}$ & $\pmb{\eta_F^I}$&$\pmb{\eta_{S_{1}}^U}$&$\pmb{\eta_{S_{2}}^U}$&$\pmb{\eta_V^U}$ & $\pmb{M_I}\,\mbox{(GeV)}$ & $\pmb{M_U}$\,\mbox{(GeV)} & $\pmb{\alpha_G^{-1}}$& $\pmb{\sin^2\theta_W}$ \\
\hline
$0$ &$0$&$0$&$0$ &$0$& $0$&$10^{13.0277}$ & $10^{14.83}$ & $40.3948$&$0.23155$ \\
\hline
$1$&$0.047$&$-2.3$&$-1$ &$-2.1$&$-2.3$& $10^{13.012}$ & $10^{15.71}$ &$36.1753$& $0.23134$ \\
\hline
$1.2$&$0.069$&$-2.3$&$-1.8$ &$-2.01$&$-2.3$& $10^{13.0036}$ & $10^{15.69}$ &$36.7186$& $0.23124$  \\
\hline
$1.4$&$0.050$&$-2.3$&$-2$ &$-2.5$&$-2.3$& $10^{13.0123}$ & $10^{15.91}$ &$33.1915$& $0.23134$ \\
\hline
$1.6$&$0.073$&$-2.3$&$-2.8$ &$-2.3$&$-2.3$& $10^{13.0035}$ & $10^{15.84}$ &$34.5268$& $0.23124$ \\

\hline \hline
\end{tabular}
\caption{Numerically estimated values for  $M_I$, $M_U$, $\alpha_G^{-1}$ and $\sin^2\theta_W$ by considering one-loop threshold effects both at $M_I$ and $M_U$  with different choices of $\eta_S^I$, $\eta_V^I$, $\eta_F^I$, $\eta_{S_{1}}^U$, $\eta_{S_{2}}^U$ and $\eta_V^U$.}.
\label{tab:threresultsMI}
\end{table} 
\end{center}
\begin{figure*}[htb!]
	\centering
	\includegraphics[width=0.45\textwidth]{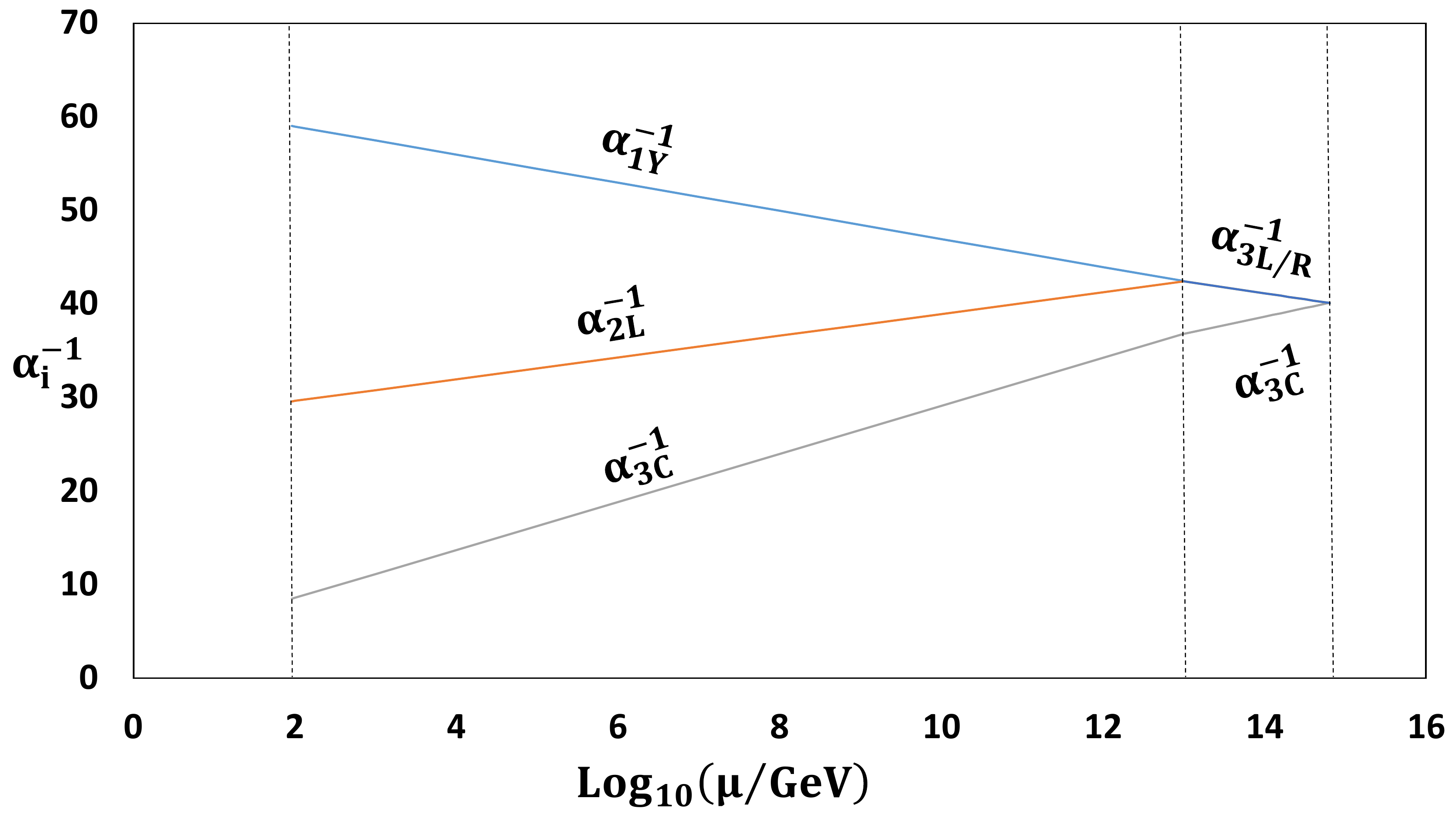}
	\includegraphics[width=0.45\textwidth]{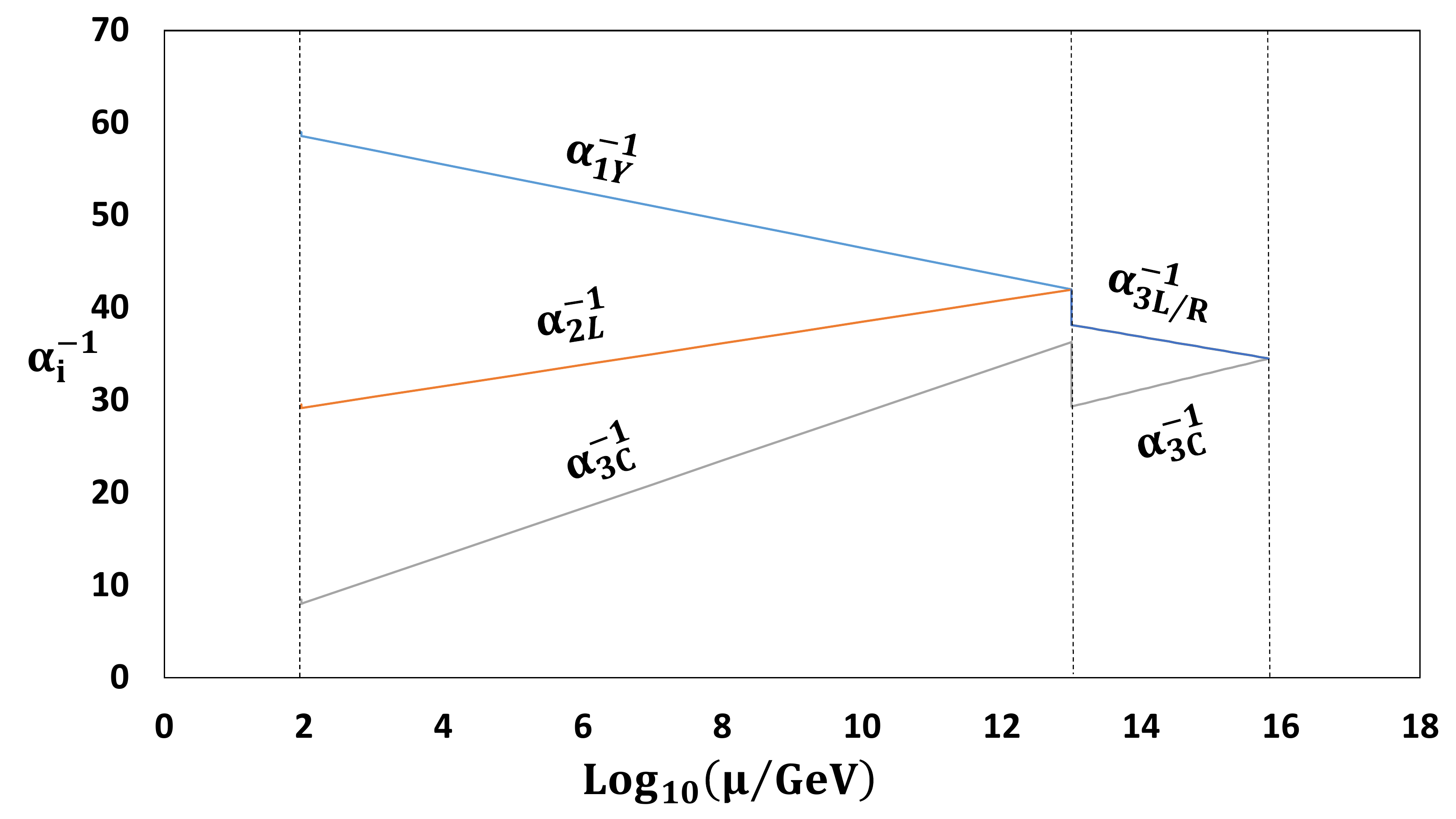}
	\caption{Gauge coupling unification plots for non-SUSY $E_6$ GUT having intermediate trinification symmetry with and without one-loop  threshold effects. The plot in the left-panel gives unification mass scale, $M_U=10^{14.81}$ GeV without threshold effects while the one in right-panel yields, $M_U=10^{15.84}$ GeV.}
	\label{fig:gamma}
\end{figure*}
\label{fig:Tri}
\end{widetext}
\vspace*{-0.95cm}
In the above table \ref{tab:threresultsMI}, the first choice corresponds to threshold effects where masses of the superheavy fields are degenerate with the symmetry breaking scale $M_U$ and $M_I$ respectively. However the above choice leads to a value of $\sin^2\theta_W=0.23155$ which is not in tune with the experimentally allowed uncertainty. With finite values of threshold parameters we can overcome the problem. The value of $M_I$ and $\sin^2\theta_W$ are affected insignificantly by threshold corrections. However, the unification mass $M_U$ increases and the inverse GUT coupling constant $\alpha_G^{-1}$ decreases with the suitable choice of the masses of the superheavy fields. Thus we achieve admissible gauge unification at $M_U$ through threshold corrections. We give the corresponding plots to show the gauge unification without threshold corrections in left-panel of Fig.\ref{fig:gamma} and with threshold corrections in right-panel of Fig.\ref{fig:gamma}. The effect enhances the unification scale so as to meet the requirements of proton decay constraint, which we discuss in the next section.

\section{Predictions on proton decay with threshold effects}
\label{sec:pdecay}
We aim to calculate the proton decay lifetime with and without one-loop threshold effects and wish to examine how the model predictions are closer or farther from the current experimental limit set by the present experiments. It is mediated mostly by the exchange of lepto-quark gauge bosons, which gives baryon and lepton number violation simultaneously. These lepto-quark gauge bosons are getting their masses through spontaneous symmetry breaking with scale VEV around mass scale $M_U$. That is the reason why one-loop GUT threshold effects are particularly important which modifies the mass scale $M_U$ and the GUT coupling constant $\pmb{\alpha_G}$ leading to important prediction for proton decay lifetime. We wish to estimate the RGE effects of this effective dimension-6 operators using Standard Model fermions till the unification scale using the relevant anomalous dimensions. 

The dimensional-6 effective operators which can induce proton decay within trinification symmetry with fermions transforming under $SU(3)_C \times SU(3)_L \times SU(3)_R$ as $\pmb{L} \equiv (1,\overline{3},3)$, $\pmb{Q} \equiv (3,3,1)$, $\pmb{Q^C} \equiv (\overline{3},1,\overline{3})$ is given below,
\begin{eqnarray}
&&\mathcal{O}^{\rm d=6}_{L}(e^C,d) \subset \left(\overline{Q^C} \gamma^\mu Q \right) 
\left(\overline{L} \gamma_\mu Q \right)  \nonumber \\
&&\mathcal{O}^{\rm d=6}_{R}(e,d^C) \subset \left(\overline{Q^C} \gamma^\mu Q \right) 
\left(\overline{Q^C} \gamma_\mu L \right) 
\end{eqnarray}

While the dimension-6 effective operator generating proton decay in terms of Standard Model fermions is as follows,
\begin{eqnarray}
&&\mathcal{O}^{\rm d=6}_{L}(e^C,d) \subset \mathbb{C}_1\epsilon^{ijk} \overline{u^C_i} \gamma^\mu u_j \, \overline{e^C} \gamma_\mu d_k   \nonumber \\
&&\mathcal{O}^{\rm d=6}_{R}(e,d^C) \subset \mathbb{C}_2
\epsilon^{ijk} \overline{u^C_i} \gamma^\mu u_j \, \overline{d^C_k} \gamma_\mu e \, 
\end{eqnarray}
with their respective Wilson coefficients $\mathbb{C}_{1,2}$. 

The master formula for the inverse of proton decay width for 
the gauge-induced dimension-6 proton decay in the chain $p \to e^+ \pi^0$ (as discussed in refs.~\cite{Babu:1992ia, Bertolini:2013vta, Kolesova:2014mfa,Parida:2016hln,Meloni:2019jcf,Ibanez:1984ni, Buras:1977yy, BhupalDev:2010he,Chakrabortty:2019fov,Babu:2015bna}) is given by
\begin{eqnarray}  
\tau_p=\pmb{\Gamma}^{-1}\left( p\rightarrow \pi^0 e^+ \right) &=&
 \frac{64 \pi f^2_\pi}{m_p} \left(\frac{M^4_U}{g^4_G} \right) \nonumber \\
 &&\hspace*{-1cm}\times \frac{1}{|\pmb{A_L}|^2 |\overline{\alpha_H}|^2 \left(1 + \mathcal{F} + \mathcal{D} \right)^2 \pmb{\mathcal{R}}} 
\label{decay-width-proton}
\end{eqnarray}
where the representative set of parameters are defined as follows,
\begin{itemize}
 \item $\pmb{A_L}$ is the long distance enhancement factor which is estimated from the RG evolution from the proton mass scale ($m_p\simeq \mbox{1\,GeV}$) to the electroweak scale ($M_Z$). 
 This enhancement factor below SM for the effective dimension-6 operator is expressed as,
$$\pmb{A_L} = \bigg[\frac{\alpha_s (\mbox{1 GeV})}{\alpha_s (m_t)}\bigg]
 ^{-\frac{4}{2 \cdot \left(-11+\frac{2}{3} \, n_f \right)}}\, ,$$
where, $n_f$ denotes the number of quark flavors at a given energy scale. Here we have neglected the effects arising from $\alpha_{2L}$ and $\alpha_{Y}$ since their contributions are very much suppressed 
as compared to the strong coupling effect $\alpha_s$. More explicitly, the enhancement factor (as derived in refs~\cite{BhupalDev:2010he,Patra:2014goa}) can be expressed as,
\begin{eqnarray}
      \hspace*{0.4cm}\pmb{A_L}=\bigg[\frac{\alpha_s (\mbox{1 GeV})}{\alpha_s (m_c)}\bigg]^{2/9} 
                               \bigg[\frac{\alpha_s (m_c)}{\alpha_s (m_b)}\bigg]^{6/25} 
                               \bigg[\frac{\alpha_s (m_b)}{\alpha_s (m_t)}\bigg]^{6/23}\simeq 1.25\,. \nonumber
\end{eqnarray}

 \item $\pmb{\mathcal{R}}$ is the renormalization factor which can be expressed as,
 
 \begin{eqnarray}
\pmb{\mathcal{R}}=\bigg[\left(\pmb{\mathcal{A}_{SL}}^2+\pmb{\mathcal{A}_{SR}}^2\right)
\left(1+ |{V_{ud}}|^2\right)^2\bigg]\, ,
 \end{eqnarray}
where, $V_{ud}=0.974$ is the $(1,1)$ element of $V_{CKM}$ mixing matrix and $\pmb{\mathcal{A}_{SL}} (\pmb{\mathcal{A}_{SR}})$ is the short-distance 
renormalization factor in the left (right) sectors derived by calculating the RGE effects from unification scale to electroweak scale. 

The short distance renormalization factor $\pmb{A_{SL(R)}}$--both for left as well as right-handed effective dimension-6 operator-- derived in the presence of all possible intermediate scales and is a model dependent factor as,
\begin{eqnarray}
&&\pmb{\mathcal{A}_{SL(R)}}= \pmb{\mathcal{A}}^{333D}_{SL(R)} \cdot \pmb{\mathcal{A}}^{213}_{SL(R)} 
\end{eqnarray}
where,  
\begin{eqnarray}
& &\pmb{\mathcal{A}}^{333D}_{SL(R)}=
          \left(\frac{\alpha^{-1}_{i} (M_I)}{\alpha^{-1}_{i} (M_{U})} \right)
          ^{ \frac{\gamma^{\prime}_{L(R)i}}{b^{\prime}_i}}, \quad 
\mbox{i=3C,3L,3R} \, ; \nonumber    \\
& &\pmb{\mathcal{A}}^{213}_{SL(R)}= \left(\frac{\alpha^{-1}_{i} (M_Z)}{\alpha^{-1}_{i} (M_{I})} \right)
          ^{ \frac{\gamma_{L(R)i}}{b_i}} \quad 
\mbox{i=2L, 1Y, 3C} \, .
\end{eqnarray}
Here $\alpha_i=g^2_i/4\pi$ is the fine structure constant for gauge group $\mathbb{G}_i$. Further $\gamma_{L(R)i}$'s  ($\gamma_{L(R)i}^{'}$'s) are the anomalous dimensions~\cite{Chakrabortty:2019fov,BhupalDev:2010he,Babu:2015bna,Patra:2014goa} given by
\begin{eqnarray*}
&&\mbox{For}\,  G_{3_{C}2_{L}1_{Y}}\,, \quad\left\{ \begin{array}{ll}
                  \gamma_L (M_Z)= \left(2, \frac{9}{4},\frac{23}{20}\right)\\
                  \gamma_R (M_Z)= \left(2, \frac{9}{4},\frac{11}{20}\right)
                 \end{array}
                 \right. 
\nonumber \\
&&\mbox{For } G_{3_{C}3_{L}3_{R}}\,,  \left\{ \begin{array}{ll}
                  \gamma_L^{\prime} (M_I)= \left(2, 2,4\right)\\
                  \gamma_R^{\prime} (M_I)= \left(2, 4,2\right)
                 \end{array}
                 \right. 
\end{eqnarray*}
and $\pmb{b_i}=(-7, -19/6, 41/10)$  ($\pmb{b^\prime_i}=(-5,7/2,7/2) $) are the one-loop beta coefficients at different stage of RGEs from $M_Z-M_{I}$ ($M_I-M_U$), respectively, presented in the Table \ref{tab:oneloopresults}.
\item Other parameters are taken from refs~\cite{Parida:2016hln,Patra:2014goa} as $\mathcal{D}=0.81$, 
$\mathcal{F}=0.47$, $f_\pi=139\, \, \mbox{MeV}$ and $m_p=938.3\, \, \mbox{MeV}$.
\end{itemize} 

Redefining $\alpha_H = \overline{\alpha}_H \left(1 + \mathcal{F} + \mathcal{D} \right) = 0.012\, \, \mbox{GeV}^3$ and 
$\pmb{A^2_R}\simeq \pmb{A^2_{L}} \left(\pmb{A^2_{SL}} + \pmb{A^2_{SR}} \right)$,  the modified expression for proton lifetime can 
be expressed as, 
\begin{eqnarray}  
\tau_{p\rightarrow \pi^0 e^+} &=&
 \frac{4}{\pi}  \left(\frac{ f^2_\pi}{m_p}\right) \left(\frac{M^4_U}{\alpha^2_G} \right)
         \frac{1}{\alpha^2_H \pmb{A^2_R} \left(1+ |{V_{ud}}|^2\right)^2}  \,
\label{lifetime-proton-modified}
\end{eqnarray}

The precision gauge coupling unification by solving RGEs for gauge coupling constants and without taking into account threshold effects give unification mass scale and inverse GUT coupling constant as,
$$M_U= 10^{14.81}\,\,\mbox{GeV and} \quad \alpha^{-1}_G=40.1059.$$

Using the numerical values of short distance renormalization factors for both the effective dimension-6 operators as $\pmb{A_{SL}}=2.46$ and $\pmb{A_{SR}}=2.34$, the  estimated proton lifetime for the present scenario (without threshold effects) is $\tau_p = 1.55 \times 10^{31}$ yrs. This prediction is well below the current 
Super-Kamiokande experiment which sets bound on the proton lifetime for $p \to e^+ \pi^0$ channel as $\tau_p (p \to e^+ \pi^0) 
> 1.6 \times 10^{34}\, \mbox{yrs}$ \cite{Miura:2016krn}  while it can be accessible to future planned experiments that can reach a bound \cite{Abe:2011ts,Yokoyama:2017mnt}
\begin{eqnarray}
& &\tau_p (p \to e^+ \pi^0) \big|_{HK, 2025} > 9.0 \times 10^{34}\, \mbox{yrs} \nonumber \\
& &\tau_p (p \to e^+ \pi^0) \big|_{HK, 2040} > 2.0 \times 10^{35}\, \mbox{yrs} 
\end{eqnarray}

It is now important to include the threshold effects both at $M_I$ and $M_U$--arising from superheavy particles (scalars, fermions and gauge bosons whose masses differ from the symmetry breaking scale)--for calculating the proton decay lifetime. 
The modified values of the unification mass scale and inverse GUT coupling constant, are given in previous section (Table \ref{tab:threresultsMI}).
Now using the values of $M_U$ and $\alpha_G^{-1}$ from Table \ref{tab:threresultsMI}, we calculate the proton decay lifetime $\tau_p$ (using eqn.(\ref{lifetime-proton-modified})). The predicted value of $\tau_p$ are given in Table \ref{tab:pdecay}. It is found that 
the estimated proton lifetime $\tau_p$ 
is consistent with the Super-Kamiokande experiments. 

\begin{table}[h]
\centering
\vspace{-2pt}
\begin{tabular}{||c|c|c||}
\hline \hline
 $\pmb{M_U}$\,\mbox{(GeV)} & $\pmb{\alpha_G^{-1}}$& $\pmb{\tau}_p$ \\
\hline
\hline
 $10^{14.81}$ & $40.1059$ 
                              & ${ 1.55 \times 10^{31} \mbox{yrs}}$  \\
\hline
 $10^{15.71}$ &$36.1753$      
                               & ${\bf 5.03 \times 10^{34} \mbox{yrs}}$  \\
\hline
 $10^{15.69}$ &$36.7186$ 
                               & ${\bf 4.31 \times 10^{34} \mbox{yrs}}$  \\
                               \hline
 $10^{15.91}$ &$33.1915$ 
                               & ${\bf 2.67 \times 10^{35} \mbox{yrs}}$ \\
\hline
 $10^{15.84}$ &$34.5268$ 
                               & ${\bf 1.51 \times 10^{35} \mbox{yrs}}$ \\
\hline \hline
\end{tabular}
\caption{Numerical estimation of proton decay lifetime $\tau_{p}$. In the last column, the bold face values for proton lifetime are in agreement with the limit set by the present Super-Kamiokande experiment.}
\label{tab:pdecay}
\end{table}
\section{Conclusion}
\label{sec:conclusion}

We have computed the threshold uncertainties for the electroweak mixing angle $\sin^2\theta_W$, intermediate mass scale $M_I$, unification mass scale $M_U$ and inverse GUT coupling constant $\alpha^{-1}_{G}$ within a class of non-supersymmetric $E_6$ Grand Unified Theory with D-parity conserving trinification symmetry $SU(3)_C\otimes SU(3)_L\otimes SU(3)_R\otimes D$. In the process, we note a crucial observation on vanishing of GUT threshold uncertainty for electroweak mixing angle $\sin^2\theta_W$ and the intermediate mass scale $M_I$. This nice feature of the model, being independent of the particle content can be generalised to all GUTs (SUSY and non-SUSY) with $G_{333D}$ intermediate symmetry. The origin behind it is primarily because of D-parity conserving trinification symmetry.\\
\\
 Coming to the quantitative effect of threshold, we see that with the conservative estimation of the unification mass scale $M_U= 10^{14.81}$~GeV and inverse GUT coupling constant $\alpha^{-1}_G=40.1059$, the predicted proton lifetime (without threshold effects) is well below the current  
Super-Kamiokande experiment which sets bound on the proton lifetime for $p \to e^+ \pi^0$ channel as $\tau_p (p \to e^+ \pi^0) 
> 1.6 \times 10^{34}\, \mbox{yrs}$ \cite{Miura:2016krn}. In order to circumvent the problem, one-loop threshold effects has been included in the model, which  yields modification in the unification mass scales, 
$M_U=10^{15.71}$~GeV ($10^{15.69}$~GeV, $10^{15.91}$~GeV, $10^{15.84}$~GeV) and $\alpha^{-1}_G$. The above estimation is with the specific choice of masses for superheavy scalars, gauge bosons and fermions which are few times heavier or lighter than the symmetry breaking mass scales $M_I$ and $M_U$. The estimated proton lifetime $\tau_p$ as, 
${\bf 5.03 \times 10^{34} \mbox{yrs}}$ (${\bf 4.31 \times 10^{34} \mbox{yrs}}$, ${\bf 2.67 \times 10^{35} \mbox{yrs}}$, ${\bf 1.51 \times 10^{35} \mbox{yrs}}$ ), respectively, is consistent with the Super-Kamiokande experiments. The threshold parameters at $M_I$ have been so choosen so as to give admissible experimental uncertainty value of electroweak mixing angle $\sin^2\theta_W$. It is observed that the threshold effects at the intermediate mass scale is very much suppressed as compared to GUT threshold effrects. The unification mass $M_U$ due to GUT threshold corrections, is controlled only by superheavy scalars, thereby increasing the predictive power of the model. This novel feature of the model is possible due to the symmetric nature of the intermediate symmetry $SU(3)_C\otimes SU(3)_L\otimes SU(3)_R\otimes D$.
Thus the present model provides an important window of opportunity for the non-supersymmetric Exceptional group $E_6$ as an attractive unification model.

\noindent
\section*{Acknowledgments}
\vspace*{-0.0cm}
Chandini Dash is grateful to the Department 
of Science and Technology, Govt. of India for INSPIRE Fellowship/2015/IF150787 in support of her research work. She acknowledges the warm hospitality provided by the IIT Bhilai  where the work has been initiated.
%

\appendix

\section{Threshold Contributions}
  The symmetry breaking channel consider here is given by 
\begin{eqnarray}
&&\pmb{E_6} \stackrel{M_U}{\longrightarrow} \mathbb{G}_{333D} 
	\stackrel{M_I}{\longrightarrow}\mathbb{G}_{321}
	\stackrel{M_Z}{\longrightarrow} \mathbb{G}_{31}
	\label{app:Model}
	\end{eqnarray}
As has been mentioned in the text, threshold effects are considered at both the symmetry breaking scales $M_U$ and $M_I$. The superheavy fields contributing to threshold are given in Table \ref{tab:app}.
Using the Table \ref{tab:app} and the general expression for the one-loop threshold effects eqn.(\ref{eq:threshold}) from the text, the one-loop threshold corrections at GUT symmetry breaking scale (or at $M_U$) are given by
\begin{eqnarray}
 \pmb{\lambda_{3C}^{U}}&=& 9+3\eta_{\phi_{2}}+3\eta_{\phi_{3}}+6\eta_{\Sigma_{5}}+9\eta_{\Sigma_{6}} +9\eta_{\Sigma_{7}}+9\eta_{\Sigma_{8}}\nonumber \\
 &&+9\eta_{\Sigma_{9}}+18\eta_{\Sigma_{10}}+18\eta_{\Sigma_{11}}+18\eta_{\Sigma_{12}}+18\eta_{\Sigma_{13}} +45\eta_{\Sigma_{14}}
  \nonumber\\
 &&+45\eta_{\Sigma_{15}}+48\eta_{\Sigma_{16}}+48\eta_{\Sigma_{17}}-\frac{189}{2}\eta_{V_{4}}-\frac{189}{2}\eta_{V_{5}}
 \label{app:L3CU} 
\end{eqnarray}
\begin{eqnarray}
\pmb{\lambda_{3L}^{U}}&=& 9+3\eta_{\phi_{2}}+6\eta_{\Sigma_{3}}+9\eta_{\Sigma_{6}}+9\eta_{\Sigma_{7}}+9\eta_{\Sigma_{8}}
 +9\eta_{\Sigma_{9}}
 \nonumber\\
 &&+45\eta_{\Sigma_{10}}
 +18\eta_{\Sigma_{11}}+45\eta_{\Sigma_{12}}
+18\eta_{\Sigma_{13}}+18\eta_{\Sigma_{14}}
  \nonumber\\
 &&+18\eta_{\Sigma_{15}}+48\eta_{\Sigma_{17}}-\frac{189}{2}\eta_{V_{4}}-\frac{189}{2}\eta_{V_{5}}
  \label{app:L3LU} 
\end{eqnarray}
\begin{eqnarray}
 \pmb{ \lambda_{3R}^{U}}&=& 9+3\eta_{\phi_{3}}+6\eta_{\Sigma_{4}}+9\eta_{\Sigma_{6}}+9\eta_{\Sigma_{7}}+9\eta_{\Sigma_{8}}
 +9\eta_{\Sigma_{9}}
  \nonumber\\
 &&+18\eta_{\Sigma_{10}}
+45\eta_{\Sigma_{11}}+18\eta_{\Sigma_{12}}+45\eta_{\Sigma_{13}}+18\eta_{\Sigma_{14}}
  \nonumber\\
 &&+18\eta_{\Sigma_{15}}+48\eta_{\Sigma_{16}}-\frac{189}{2}\eta_{V_{4}}-\frac{189}{2}\eta_{V_{5}}
   \label{app:L3RU}
\end{eqnarray}

Similarly, one-loop threshold contributions $\pmb{\lambda_i^I}$ are given by,       
\begin{eqnarray}
 \pmb{\lambda_{3C}^I}&=&4\eta_{D}+4\eta_{D^{C}} \nonumber \\
\pmb{\lambda_{2L}^I}&=&1+\eta_{\phi_{12}}+\eta_{\phi_{13}}+4\eta_{\psi}+4\eta_{\psi^{C}}-\frac{21}{2}\eta_{V_{22}}-\frac{21}{2}\eta_{V_{23}} \nonumber \\
\pmb{\lambda_{Y}^I}&=&3+\frac{3}{5}\eta_{\phi_{12}}+\frac{3}{5}\eta_{\phi_{13}}+\frac{6}{5}\eta_{\phi_{14}}+\frac{8}{5}\eta_{D}+\frac{8}{5}\eta_{D^{C}}
\nonumber \\
&&+\frac{12}{5}\eta_{\psi}+\frac{12}{5}\eta_{\psi^{C}}-\frac{63}{10}\eta_ {V_{22}}-\frac{63}{10}\eta_ {V_{23}}-\frac{63}{5}\eta_ {V_{31}}\nonumber\\
&&-\frac{63}{5}\eta_ {V_{34}}-\frac{63}{5}\eta_ {V_{35}}-\frac{63}{5}\eta_ {V_{36}}
\label{app:LI}
\end{eqnarray}
 We then follow the assumptions mentioned in the text regarding the masses of the superheavy particles, we obtain the threshold corrections as,
 \begin{eqnarray}
  \pmb{\lambda_{3C}^{U}}&=&9+6\eta_{S_{1}}^{U}+300\eta_{S_{2}}^{U}-189\eta_V^U\nonumber\\
  \pmb{\lambda_{3L}^{U}}&=&9+3\eta_{S_{1}}^{U}+252\eta_{S_{2}}^{U}-189\eta_V^U\nonumber\\
  \pmb{ \lambda_{3R}^{U}}&=&9+3\eta_{S_{1}}^{U}+252\eta_{S_{2}}^{U}-189\eta_V^U
 \end{eqnarray}
and 
\begin{eqnarray}
&&\pmb{\lambda_{3C}^I}= 8\eta_F^I\,, \quad  \pmb{\lambda_{2L}^I}=1+2\eta_S^I+8\eta_F^I-21\eta_V^I\,, \nonumber\\
&&\pmb{\lambda_{Y}^I}=3+\frac{12}{5}\eta_S^I+8\eta_F^I-63\eta_V^I
 \label{app:LI-Th}
\end{eqnarray}

where $\eta_{S_{1}}^{U}= \ln \frac{M_{S_{1}}^{U}}{M_U}$, $\eta_{S_{2}}^{U}= \ln \frac{M_{S_{2}}^{U}}{M_U}$, $\eta_V^U= \ln \frac{M_V^U}{M_U}$ and $\eta_S^I= \ln \frac{M_S^{I}}{M_I}$, $\eta_F^I=\ln \frac{M_F^{I}}{M_I}$, $\eta_V^I= \ln \frac{M_V^{I}}{M_I}$. 
\newpage
\begin{widetext}
\begin{center}
\begin{table}[htb!]
\begin{tabular}{||c||c||c||c||}
\hline \hline
\pmb{Fields} & $\pmb{E_6}$ & $\pmb{G_{3_{C}3_{L}3_{R}D}}$(Fields at $M_{U}$)& $\pmb{G_{3_{C}2_{L}1_{Y}}}$(Fields at $M_{I}$) \\
\hline
Fermion & $27_F$ & $ \begin{array}{clcr}
                                \pmb{L(1,\overline{3},3)}\\ 
                               \pmb{ Q(3,3,1)}\\\pmb{Q^{C}(\overline{3},1,\overline{3})} \end{array} $
& $\begin{array}{clcr}
                                \pmb{Q(3,2,\frac{1}{6})}, 
                               \pmb{ u^{C}(\overline{3},1,-\frac{2}{3})},\pmb{d^{C}(\overline{3},1,\frac{1}{3})} \\\pmb{l(1,2,-\frac{1}{2})},\pmb{e^{C}(1,1,1)}\\
                               D(3,1,-\frac{1}{3}), D^{C}(\overline{3},1,\frac{1}{3})\\\psi(1,2,-\frac{1}{2}),\psi^{C}(1,2,\frac{1}{2})\\
                               \rho(1,1,0),\nu^{C}(1,1,0)\end{array}$ 
                              \\
\hline
Scalar & $27_H$ & $ \begin{array}{clcr}
                                \pmb{\phi_{1}(1,\overline{3},3)}\\ 
                                \phi_{2}(3,3,1),\phi_{3}(\overline{3},1,\overline{3}) \end{array} $  & $\begin{array}{clcr}\pmb{\phi(1,2,-\frac{1}{2})}\\\phi_{11}(1,1,0),\phi_{12}(1,2,-\frac{1}{2}),\phi_{13}(1,2,\frac{1}{2})\\\phi_{14}(1,1,1),\phi_{15}(1,1,0)\end{array}$ 
                                \\
\hline
Scalar & $650_{H}$ & $\begin{array}{clcr}
                        \pmb{\Sigma_{0}(1,1,1)},\pmb{\Sigma_{1}(1,8,8)}\\
                        \Sigma_{2}(1,1,1),\Sigma_{3}(1,8,1),\Sigma_{4}(1,1,8),\Sigma_{5}(8,1,1)\\\Sigma_{6}(\overline{3},3,3),\Sigma_{7}(\overline{3},3,3),\Sigma_{8}(3,\overline{3},\overline{3})\\\Sigma_{9}(3,\overline{3},\overline{3}),\Sigma_{10}(3,6,\overline{3}),\Sigma_{11}(3,\overline{3},6)\\\Sigma_{12}(\overline{3},\overline{6},3),\Sigma_{13}(\overline{3},3,\overline{6}),\Sigma_{14}(\overline{6},\overline{3},\overline{3})\\\Sigma_{15}(6,3,3),\Sigma_{16}(8,1,8),\Sigma_{17}(8,8,1)
                       \end{array}
$  &  \\
\hline 
Gauge Boson & $78_V$ & $\begin{array}{clcr}
                        \pmb{V_{1}(8,1,1)},\pmb{V_{2}(1,8,1)},\pmb{V_{3}(1,1,8)}\\
                        V_{4}(\overline{3},3,3),V_{5}(3
                        ,\overline{3},\overline{3})
                       \end{array}$  & $\begin{array}{clcr}
                        \pmb{V_{10}(8,1,0)},\pmb{V_{20}(1,3,0)},\pmb{V_{30}(1,1,0)}\\V_{21}(1,1,0),V_{22}
                        (1,2,\frac{1}{2}),V_{23}(1,2,-\frac{1}{2})\\V_{31}(1,1,1),V_{32}(1,1,0),V_{33}(1,1,0)\\V_{34}(1,1,-1),V_{35}(1,1,1),V_{36}(1,1,-1),V_{37}(1,1,0)\end{array}$ 
                       \\
\hline 
\hline
\end{tabular}
\caption{The superheavy scalars, fermions and gauge bosons at different symmetry breaking scales arising from $E_6$ representations $27_F, 27_H, 650_H, 78_V$. The superheavy fields denoted in normal text transforming under trinification symmetry are presented in third column while for SM symmetry in fourth column. Here the light fields (scalars, fermions and gauge bosons) denoted in bold face in third and fourth column are not contributing to one-loop threshold effects but take part in the RG evolution of gauge couplings.}
\label{tab:app}
\end{table}
\end{center}
\end{widetext}
%

\bibliographystyle{utcaps_mod}
\bibliography{E62}
\end{document}